\input phyzzx
\overfullrule=0pt
\def\tr{\mathop{\rm tr}\nolimits}
\def\sqr#1#2{{\vcenter{\hrule height.#2pt
      \hbox{\vrule width.#2pt height#1pt \kern#1pt
          \vrule width.#2pt}
      \hrule height.#2pt}}}

\def\square{{\mathchoice{\sqr84}{\sqr84}{\sqr{5.0}3}{\sqr{3.5}3}}}
\def\dA{\mathop\square\nolimits}
\def\Mto{{\buildrel{M\rightarrow\infty}\over=}}
\REF\BAR{%
W. A. Bardeen,
Phys.\ Rev.\ {\bf 184} (1969) 1848.}
\REF\FUJ{%
K. Fujikawa,
Phys.\ Rev.\ Lett.\ {\bf 42} (1979) 1195; {\bf 44} (1980) 1733;
Phys.\ Rev.\ {\bf D21} (1980) 2848; {\bf D22} (1980) 1499~(E);
{\bf D23} (1981) 2262.}
\REF\WES{%
J. Wess and B. Zumino,
Phys.\ Lett.\ {\bf 37B} (1971) 95.}
\REF\BARD{%
W. A. Bardeen and B. Zumino,
Nucl.\ Phys.\ {\bf B244} (1984) 421.}
\REF\ZUM{%
B. Zumino, Y. S. Wu and A. Zee,
Nucl.\ Phys.\ {\bf B239} (1984) 477.\nextline
B. Zumino,
in {\sl Relativity, Groups and Topology II}, eds.\ B. S. De Witt and
R. Stora, (North-Holland, Amsterdam, 1984).}
\REF\BAU{%
L. Baulieu,
Nucl.\ Phys.\ {\bf B241} (1984) 557;
in {\sl Progress in Gauge Field Theory}, eds.\ H. Lehmann et.\ al.,
NATO ASI Series B, Physics, Vol.~115 (Plenum, New York, 1984).}
\REF\STO{%
R. Stora,
in {\sl Progress in Gauge Field Theory}, eds.\ H. Lehmann et.\ al.,
NATO ASI Series B, Physics, Vol.~115 (Plenum, New York, 1984).}
\REF\ALV{%
L. Alvarez-Gaum\'e and P. Ginsparg,
Nucl.\ Phys.\ {\bf B243} (1984) 449.}
\REF\BAN{%
H. Banerjee, R. Banerjee and P. Mitra,
Z. Phys.\ {\bf C32} (1986) 445.}
\REF\LEU{%
H. Leutwyler,
Phys.\ Lett.\ {\bf B152} (1985) 78.}
\REF\OKU{%
K. Okuyama and H. Suzuki,
Phys.\ Rev.\ {\bf D56} (1997) 6829.\nextline
H. Suzuki,
hep-lat/9901012, to appear in Prog.\ Theor.\ Phys.}
\REF\PIG{%
O. Piguet and K. Sibold,
Nucl.\ Phys.\ {\bf B247} (1984) 484.}
\REF\CLA{%
T. E. Clark and S. Love,
Phys.\ Lett.\ {\bf 138B} (1984) 289.}
\REF\NIE{%
N. K. Nielsen,
Nucl.\ Phys.\ {\bf B244} (1984) 499.}
\REF\GUA{%
E. Guadagnini, K. Konishi and M. Mintchev,
Phys.\ Lett.\ {\bf 157B} (1985) 37.}
\REF\HAR{%
K. Harada and K. Shizuya,
Phys.\ Lett.\ {\bf 162B} (1985) 322.}
\REF\NEM{%
D. Nemeschansky and R. Rohm,
Nucl.\ Phys.\ {\bf B249} (1985) 157.}
\REF\GIR{%
G. Girardi, R. Grimm and R. Stora,
Phys.\ Lett.\ {\bf 156B} (1985) 203.}
\REF\BON{%
L. Bonora, P. Pasti and M. Tonin,
Phys.\ Lett.\ {\bf 156B} (1985) 341; Nucl.\ Phys.\ {\bf B261} (1985)
249.}
\REF\GAR{%
R. Garreis, M. Scholl and J. Wess,
Z. Phys.\ {\bf C28} (1985) 623.}
\REF\PER{%
M. Pernici and F. Riva,
Nucl.\ Phys.\ {\bf B267} (1986) 61.}
\REF\KRI{%
V. K. Krivoshchekov, P. B. Medvedev and L. O. Chekhov,
Theor.\ Math.\ Phys.\ {\bf 68} (1987) 796.}
\REF\MCA{%
I. N. McArthur and H. Osborn,
Nucl.\ Phys.\ {\bf B268} (1986) 573.}
\REF\MAR{%
M. Marinkovi\'c,
Nucl.\ Phys.\ {\bf B366} (1991) 74.}
\REF\WESS{%
J. Wess and J. Bagger,
``Supersymmetry and Supergravity,''
(Princeton University Press, Princeton, 1992).}
\REF\HAY{%
T. Hayashi, Y. Ohshima, K. Okuyama and H. Suzuki,
Prog.\ Theor.\ Phys.\ {\bf 100} (1998) 627.}
\REF\KON{%
K. Konishi and K. Shizuya,
Nuovo Cim.\ {\bf 90A} (1985) 111.}
\REF\HAYA{%
T. Hayashi, Y. Ohshima, K. Okuyama and H. Suzuki,
Prog.\ Theor.\ Phys.\ {\bf 100} (1998) 1033.}
\REF\FRO{%
S. A. Frolov and A. A. Slavnov,
Phys.\ Lett.\ {\bf B309} (1993) 344.}
\REF\NAR{%
R. Narayanan and H. Neuberger,
Phys.\ Lett.\ {\bf B302} (1993) 62.}
\REF\NIEM{%
A. J. Niemi and G. W. Semenoff,
Phys.\ Rev.\ Lett.\ {\bf 55} (1985), 927; {\bf 55} (1985), 2627 (E).}
\REF\BAL{%
R. D. Ball, Phys.\ Rep.\ {\bf 182} (1989), 1.}
\REF\NEU{%
H. Neuberger,
Phys.\ Rev.\ {\bf D59} (1999) 085006.}
\REF\ITO{%
H. Itoyama, V. P. Nair and H. Ren,
Nucl.\ Phys.\ {\bf B262} (1985) 317.}
\REF\GUAD{%
E. Guadagnini and M. Mintchev,
Nucl.\ Phys.\ {\bf B269} (1986) 543.}
%
\Pubnum={IU-MSTP/33; hep-th/9904096}
\date={April 1999}
\titlepage
\title{Remark on the Consistent Gauge Anomaly\break
in Supersymmetric Theories}
\author{%
Yoshihisa Ohshima,\foot{%
E-mail address: ohshima@mito.ipc.ibaraki.ac.jp}
Kiyoshi Okuyama\foot{%
E-mail address: okuyama@mito.ipc.ibaraki.ac.jp}
Hiroshi Suzuki\foot{%
E-mail address: hsuzuki@mito.ipc.ibaraki.ac.jp}
and Hirofumi Yasuta\foot{%
E-mail address: yasuta@mito.ipc.ibaraki.ac.jp}}
\address{%
Department of Physics, Ibaraki University, Mito 310-8512, Japan}
\abstract{%
We present a direct field theoretical calculation of the consistent
gauge anomaly in the superfield formalism, on the basis of a
definition of the effective action through the covariant gauge
current. The scheme is conceptually and technically simple and the
gauge covariance in intermediate steps reduces calculational labors
considerably. The resultant superfield anomaly, being proportional to
the anomaly~$d^{abc}=\tr T^a\{T^b,T^c\}$, is minimal without
supplementing any counterterms. Our anomaly coincides with the
anomaly obtained by Marinkovi\'c as the solution of the Wess-Zumino
consistency condition.}
\endpage
The consistent gauge anomaly~[\BAR] might be conceptually more
natural than the covariant gauge anomaly~[\FUJ], as it is defined as
gauge non-invariance of the effective action of the chiral
fermion~[\WES,\BARD]. The consistent anomaly is important because it
provides information on the Wess-Zumino Lagrangian~[\WES]. To find an
explicit form of the consistent anomaly, one may appeal to the
algebraic-geometrical technique~[\ZUM-\ALV] or directly perform a
field theoretical calculation with, say, the Pauli-Villars
regularization. As is well-known, however, both approaches can be
cumbersome for a theory in higher dimensions, or for a theory with a
complicated gauge transformation. In particular, in the field
theoretical calculation, gauge non-invariant normal terms
(fake anomalies) generally appear. Then, to extract the intrinsic
anomaly, one has to find suitable local counterterms to eliminate
these normal terms.

The covariant gauge anomaly, on the other hand, has the quite
restricted possible form due to the gauge covariance. The necessary
calculational labors are consequently considerably less. Therefore, a
practically useful calculational scheme might be formulated by
relating the consistent anomaly with the covariant anomaly (or with a
certain gauge covariant expression). In~Ref.~[\BAN], Banerjee,
Banerjee and Mitra gave a field theoretical prescription which
provides this kind of calculational scheme. This prescription leads
to basically equivalent consequences as the result due to Bardeen
and Zumino~[\BARD], and that due to Leutwyler~[\LEU]. However, the
prescription of~Ref.~[\BAN] is more straightforward and
flexible.\foot{%
It has the application even in chiral gauge theories on the
lattice~[\OKU].}

In this letter, we give a {\it direct\/} field theoretical
calculation of the consistent gauge anomaly in supersymmetric
theories, on the basis of the prescription of~Ref.~[\BAN]. Generally,
the treatment of the consistent anomaly with the superfield
formalism~[\PIG--\MAR] is quite complicated, because the gauge
transformation is highly non-linear and because the gauge superfield
has no mass dimension (i.e., an arbitrary function of the gauge
superfield is a candidate of the counterterm). The advantage of our
treatment in this letter is that the minimal superfield anomaly,
being proportional to the anomaly~$d^{abc}=\tr T^a\{T^b,T^c\}$, is
directly obtained. This minimal-ness is guaranteed by the basic
property of the prescription of~Ref.~[\BAN]. Naturally, the resultant
anomaly coincides with that due to Marinkovi\'c~[\MAR], who applied
the technique of~Ref.~[\BARD] to this problem. Also our expressions
below have some similarities with that of the work
by~McArthur and~Osborn~[\MCA], in which the formulation
of~Ref.~[\LEU] was generalized to supersymmetric theories.
Nevertheless, it seems worthwhile to report on our field theoretical
calculation, because of simplicity of the basic idea and the
treatment.

We consider the massless chiral superfield~$\Phi$ coupled to the
external gauge superfield~$V=V^aT^a$ ($T^a$~is the representation of
the gauge group to which $\Phi$ belongs). The classical action is
given by\foot{%
We basically follow the notational conventions of~Ref.~[\WESS]. Our
particular conventions and useful identities are summarized in the
Appendix~A.}
$$
   S=\int d^8z\,\Phi^\dagger e^V\Phi.
\eqn\one
$$
Following the prescription of~Ref.~[\BAN], we define the effective
action~${\mit\Gamma}[V]$ as follows. We first introduce an auxiliary
gauge coupling parameter\foot{%
As one would anticipate, this parameter~$g$ becomes the integration
variable appearing in the homotopy formula~[\ZUM-\ALV,\BARD].} $g$
by~$V\to gV$. Then we may differentiate the effective action with
respect to the parameter~$g$ and integrate it over this parameter.
Noting that the $g$-dependences arise only through the
combination~$gV$ and the original effective action is given by the
value at~$g=1$, we have the following formal expression of the
effective action
$$
\eqalign{
   {\mit\Gamma}[V]
   &=\int_0^1dg\,\int d^8z\,V^a(z)
   {\delta{\mit\Gamma}[gV]\over\delta gV^a(z)}
\cr
   &=\int_0^1dg\,\int d^8z\,V^a(z)
   \VEV{{\delta S\over\delta V^a(z)}}_{V\to gV}.
\cr
}
\eqn\two
$$
Here the indication~$V\to gV$ implies that all $V$-dependences
involved are replaced by~$gV$. The representation~\two\ is yet
formal, because the regularization of the gauge
current~$\VEV{\delta S/\delta V^a(z)}$ has to be specified. The
crucial point of the prescription of~Ref.~[\BAN] is to adopt the
{\it covariant\/} gauge current as the gauge current. Thus we
introduce the proper time cutoff to regularize the gauge current
in a gauge covariant manner
$$
\eqalign{
   \VEV{{\delta S\over\delta V^a(z)}}
   &=\lim_{z'\to z}\tr{\partial e^V\over\partial V^a}
   \VEV{{\rm T}^*\Phi(z)\Phi^\dagger(z')}
\cr
   &\equiv-{i\over16}\Tr
   e^{-V}{\delta e^V\over\delta V^a(z)}
   \overline D^2\int_{1/M^2}^\infty dt\,
   e^{\dA_+t}\nabla^2
\cr
   &={i\over16}\Tr
   e^{-V}{\delta e^V\over\delta V^a(z)}
   \overline D^2e^{\dA_+/M^2}{1\over\dA_+}\nabla^2,
\cr
}
\eqn\three
$$
where the trace~$\Tr$ is taken with the full superspace
measure~$d^8z$ and $M$~denotes the cutoff mass parameter.\foot{%
One can generalize the regulator~$e^{\dA_+/M^2}$ in the last line
as~$f(-\dA_+/M^2)$ where $f(x)$~is an arbitrary rapidly decreasing
function with~$f(0)=1$. The result with~$f(-\dA_+/M^2)$ is given by
working with~$e^{\dA_+p/M^2}$ and then by
multiplying~$\int_0^\infty dp\,g(p)$; $g(p)$~is the inverse
Laplace transformation of~$f(x)$,
$f(x)=\int_0^\infty dp\,g(p)e^{-px}$. Our results in~$M\to\infty$ are
independent of~$M$ and thus of~$p$. Therefore, all the results become
independent of the choice of the regulator function~$f(x)$
in the $M\to\infty$~limit because $\int_0^\infty dp\,g(p)=f(0)=1$.}
In writing this expression, we have used the formal expression of the
propagator of the chiral superfield in presence of the external gauge
superfield $\VEV{{\rm T}^*\Phi(z)\Phi^\dagger(z')}=%
i\overline D^2{1\over\dA_+}\nabla^2e^{-V}\delta(z-z')/16$. (For the
derivation of the propagator, see, for example, Ref.~[\HAY].) Note
that all the derivatives~$\nabla_\alpha$, $\overline D_{\dot\alpha}$
and~$\nabla_m$ transform
as~$\nabla'=e^{-i\Lambda}\nabla e^{i\Lambda}$ under the gauge
transformation~$e^{V'}=e^{-i\Lambda^\dagger}e^Ve^{i\Lambda}$~[\WESS]
and  thus these are gauge covariant objects. Due to the gauge
covariant definition~\three, the gauge current transforms covariantly
under the gauge transformation,
$$
\eqalign{
   \VEV{{\delta S\over\delta V^a(z)}}'
   &={i\over16}\Tr
   e^{-V}{\delta e^V\over\delta V^{\prime a}(z)}
   \overline D^2e^{\dA_+/M^2}{1\over\dA_+}\nabla^2
\cr
   &=\int d^8z'\,
   {\delta V^b(z')\over\delta V^{\prime a}(z)}
   \VEV{{\delta S\over\delta V^b(z')}},
\cr
}
\eqn\four
$$
as is formally expected.

The definition of the effective action~\two\ through the covariant
current~\three\ is perfectly legitimate. For UV convergent diagrams,
it is equivalent to any conventional definition. For UV diverging
diagrams, it may give a different value from the conventional
definition but only by an amount expressed by local counterterms,
because Eq.~\three\ reduces to the conventional gauge current in the
$M\to\infty$~limit. As we will see below, the consistent anomaly
derived from the effective action~\two\ with~Eq.~\three\ is directly
related to the {\it covariant\/} anomaly. Since the covariant
anomaly~[\KON,\MCA] is proportional to the anomaly~$d^{abc}$, the
consistent anomaly thus obtained is also proportional to~$d^{abc}$.
This implies that, when the gauge representation of the chiral
multiplet is anomaly-free, i.e., when~$d^{abc}=0$, the regularized
effective action~\two\ with~Eq.~\three\ automatically restores the
gauge invariance {\it without\/} supplementing any counterterms. In
this sense, a breaking of the gauge symmetry is kept to be
minimal with the present prescription~[\HAYA].\foot{%
For anomaly-free cases, the prescription is equivalent~[\HAYA] to the
generalized Pauli-Villars regularization introduced
in~Refs.~[\FRO,\NAR]. Since this is a Lagrangian level regularization,
and the corresponding Hamiltonian is Hermitian, the S-matrix is
manifestly unitary. (In the $M\to\infty$~limit, negative norm
regulators cannot contribute to the out-state of the physical
S-matrix.)} The same mechanism works also when one starts with the
covariant current and then adds minimal corrections for ensuring the
integrability of the whole current~[\LEU,\MCA]. Our treatment is,
however, more straightforward as it directly defines the effective
action. Note that the prescription~\two\ and~\three\ is quite
different from the direct proper time regularization of the effective
action~[\NIE,\GUA].

Now, from Eq.~\two, we can read off a variation of the effective
action under the infinitesimal gauge transformation~[\WESS]
$$
\eqalign{
   &\delta_\Lambda e^V=e^Vi\Lambda-i\Lambda^\dagger e^V,
\cr
   &\delta_\Lambda V=
   i{\cal L}_{V/2}\cdot\Bigl[(\Lambda+\Lambda^\dagger)
   +\coth({\cal L}_{V/2})\cdot(\Lambda-\Lambda^\dagger)\Bigr],
\cr
}
\eqn\five
$$
as
$$
\eqalign{
   \delta_\Lambda{\mit\Gamma}[V]
   &=\int_0^1dg\,\int d^8z\,\delta_\Lambda V^a(z)
   \VEV{{\delta S\over\delta V^a(z)}}_{V\to gV}
\cr
   &\quad
   +\int_0^1dg\,\int d^8z\,\int d^8z'\,V^a(z)
   \delta_\Lambda V^b(z'){\delta\over\delta V^b(z')}
   \VEV{{\delta S\over\delta V^a(z)}}_{V\to gV}.
\cr
}
\eqn\six
$$
We then insert~$dg/dg=1$ into the first term and perform the
integration by parts with respect to~$g$. By noting again that the
$g$-dependences arise only through the combination~$gV$, we have the
following representation of the consistent anomaly
$$
\eqalign{
   &\delta_\Lambda{\mit\Gamma}[V]
   =\int d^8z\,\delta_\Lambda V^a(z)
   \VEV{{\delta S\over\delta V^a(z)}}
\cr
   &\quad+\int_0^1dg\,\int d^8z'\,\delta_\Lambda V^b(z')
\cr
   &\qquad\quad\times
   \int d^8z\,\biggl\{V^a(z)
   \biggl[{\delta\over\delta V^b(z')}
   \VEV{{\delta S\over\delta V^a(z)}}
   -{\delta\over\delta V^a(z)}
   \VEV{{\delta S\over\delta V^b(z')}}\biggr]\biggr\}_{V\to gV}.
\cr
}
\eqn\seven
$$
This anomaly must satisfy the Wess-Zumino consistency condition
because it {\it is\/} a variation of the effective action~\two.
Eq.~\seven\ shows that the consistent anomaly consists of two pieces:
The first piece is the covariant gauge anomaly~[\KON,\MCA] that is
obtained from~Eqs.~\three\ and~\five\ as
$$
\eqalign{
   &\int d^8z\,\delta_\Lambda V^a(z)\VEV{{\delta S\over\delta V^a(z)}}
\cr
   &=i\int d^8z\,\lim_{z'\to z}\tr i\Lambda
   e^{\dA_+/M^2}{1\over\dA_+}
   {1\over16}\overline D^2\nabla^2
   \delta(z-z')+{\rm h.c.}
\cr
   &=-{i\over4}\int d^6z\,\lim_{z'\to z}\tr i\Lambda
   e^{\dA_+/M^2}\overline D^2
   \delta(z-z')+{\rm h.c.}
\cr
   &\Mto-{1\over64\pi^2}\int d^6z\,\tr i\Lambda
   W^\alpha W_\alpha
   +{1\over64\pi^2}\int d^6\overline z\,\tr e^{-V}i\Lambda^\dagger e^V
   \overline W_{\dot\alpha}^\prime\overline W^{\prime\dot\alpha},
\cr
}
\eqn\eight
$$
where we have noted that~$\int d^8z=\int d^6z\,(-\overline D^2/4)$ and
the gauge parameter is chiral $\overline D_{\dot\alpha}\Lambda=0$. We
have also used the identity~(A.3). For the actual calculation of the
third line in plane wave basis, see~Ref.~[\KON], or the Appendix
of~Ref.~[\HAY]. The calculation is quite simple due to the gauge
covariance. Obviously the covariant anomaly~\eight\ is proportional
to the anomaly~$d^{abc}$, because $\Lambda$, $W_\alpha$,
$\Lambda^\dagger$, and~$\overline W_{\dot\alpha}^\prime$ are Lie
algebra valued.

The second piece in~Eq.~\seven, on the other hand, provides a
difference between the consistent anomaly and the covariant
anomaly~[\BARD]. The difference is expressed by the functional
rotation of the covariant gauge current
$$
   {\delta\over\delta V^b(z')}
   \VEV{{\delta S\over\delta V^a(z)}}
   -{\delta\over\delta V^a(z)}
   \VEV{{\delta S\over\delta V^b(z')}}.
\eqn\nine
$$
The importance of this functional rotation has been noticed in
various context~[\ALV--\LEU,\NIEM,\MCA,\BAL,\OKU,\NEU]. The gauge
covariance~\four\ implies that the functional rotation~\nine\
possesses the following property:
$$
\displaylines{
   \quad
   \int d^8z\,\delta_\Lambda V^a(z)
   \biggl[{\delta\over\delta V^b(z')}
   \VEV{{\delta S\over\delta V^a(z)}}
   -{\delta\over\delta V^a(z)}
   \VEV{{\delta S\over\delta V^b(z')}}\biggr]
   \hfill
\cr
   \hfill{}={\delta\over\delta V^b(z')}
   \int d^8z\,\delta_\Lambda V^a(z)
   \VEV{{\delta S\over\delta V^a(z)}}.
   \quad\eqnalign{\ten}
\cr
}
$$
The right hand side is nothing but the covariant anomaly~\eight.
Quite interestingly, the functional rotation~\nine\ is a
{\it local\/} functional of the gauge superfield, being proportional
to (a derivative of) the delta function~$\delta(z-z')$. We will see
this shortly. Therefore, Eq.~\ten\ implies that, when the covariant
anomaly vanishes, i.e., when~$d^{abc}=0$, the functional rotation
vanishes and consequently our consistent anomaly~\seven\ entirely
vanishes. Thus the minimal-ness of the anomaly is guaranteed by
construction; this is the advantage of the prescription~\two\
and~\three. In non-super\-symmetric gauge theories, one can obtain
the functional rotation~\nine\ by solving the relation corresponding
to~Eq.~\ten~[\BAN]. Instead, one may directly evaluate Eq.~\nine, as
was performed in~Ref.~[\LEU] for non-super\-symmetric gauge theories.
Here we adopt the latter approach that seems much simpler for the
present supersymmetric case. This approach was also adopted
in~Ref.~[\MCA].

To evaluate the functional rotation~\nine, we consider
$$
   \delta_1\VEV{\delta_2S}-\delta_2\VEV{\delta_1S},
\eqn\eleven
$$
where $\delta_1$ and~$\delta_2$ are arbitrary variations of the
gauge superfield~$V$, being independent of~$V$ itself.
Here, in the same way as the gauge current~\three, the composite
operator~$\VEV{\delta S}$ is regularized in a gauge covariant manner
$$
\eqalign{
   \VEV{\delta S}
   &\equiv-{i\over16}\Tr\Delta\overline D^2\int_{1/M^2}^\infty dt\,
   e^{\dA_+t}\nabla^2
\cr
   &=-{i\over16}\Tr\Delta\overline D^2\int_{1/M^2}^\infty dt\,
   e^{\nabla^2\overline D^2t/16}\nabla^2,
\cr
}
\eqn\twelve
$$
with the notation~$\Delta\equiv e^{-V}\delta e^V$. By noting
relations $\delta\nabla^2=\bigl[\nabla^2,\Delta\bigr]$ and
$\delta_1\Delta_2=-\Delta_1\Delta_2%
+\hbox{(symmetric on $1\leftrightarrow2$)}$, we have
$$
\eqalign{
   &\delta_1\VEV{\delta_2S}-\delta_2\VEV{\delta_1S}
\cr
   &={i\over16}\int_{1/M^2}^\infty dt\,\Tr\Biggl\{\int_0^tds\,
   \biggl[\Delta_2\overline D^2e^{\nabla^2\overline D^2s/16}
   \Delta_1{\partial\over\partial t}e^{\nabla^2\overline D^2(t-s)/16}
   \nabla^2\biggr]
\cr
   &\qquad\qquad\qquad\qquad\qquad\qquad\qquad\quad
   +\Delta_2\overline D^2e^{\nabla^2\overline D^2t/16}\Delta_1
   \nabla^2\Biggr\}-(1\leftrightarrow2)
\cr
   &=-{i\over16}\int d^8z\,\int_0^1d\beta\,
   {1\over M^2}\lim_{z'\to z}\tr
   \Delta_2e^{\beta\dA_+/M^2}\overline D^2
   \Delta_1e^{(1-\beta)\dA_-/M^2}\nabla^2
   \delta(z-z')
\cr
   &\qquad\qquad\qquad\qquad\qquad\qquad\qquad\qquad\qquad\qquad
   \qquad\qquad\qquad\quad
   -(1\leftrightarrow2).
\cr
}
\eqn\thirteen
$$
In writing the last expression, we have used the identities~(A.3)
and~(A.4). Note that, while originally the proper time in the gauge
current~\twelve\ is belonging to the IR region~$1/M^2\leq t<\infty$,
only the UV region~$0\leq\beta/M^2\leq1/M^2$
(or $0\leq(1-\beta)/M^2\leq1/M^2$) is contained in the
combination~\thirteen. Thus the functional rotation~\eleven\
or~\nine\ is a local quantity like the anomaly itself. Thanks to the
gauge covariance, the evaluation of the last expression in
plane wave basis is again simple as is shown in Appendix~B. We have
$$
\eqalign{
   &\delta_1\VEV{\delta_2S}-\delta_2\VEV{\delta_1S}
\cr
   &\Mto{1\over64\pi^2}\int d^8z\,
   \tr\Delta_1\Bigl(
   \bigl[{\cal D}^\alpha\Delta_2,W_\alpha\bigr]
   +\bigl[\overline D_{\dot\alpha}\Delta_2,
          \overline W^{\prime\dot\alpha}\bigr]
   +\bigl\{\Delta_2,{\cal D}^\alpha W_\alpha\bigr\}\Bigr).
\cr
}
\eqn\fourteen
$$
In spite of the asymmetric appearances of~$1$ and~$2$ in this
expression, one can confirm by using the reality constraint~(A.2)
that this is actually odd under the exchange~$1\leftrightarrow2$.
{}From Eq.~\fourteen, we can read off the left hand side of~Eq.~\ten:
$$
\eqalign{
   &\int d^8z\,\delta_\Lambda V^a(z)
   \biggl[{\delta\over\delta V^b(z')}
   \VEV{{\delta S\over\delta V^a(z)}}
   -{\delta\over\delta V^a(z)}
   \VEV{{\delta S\over\delta V^b(z')}}\biggr]
\cr
   &\Mto{1\over64\pi^2}\int d^8z\,
   \int_0^1d\beta
\cr
   &\qquad\quad\times
   \tr e^{-\beta V}T^b\delta(z-z')e^{\beta V}
   \Bigl({\cal D}^\alpha\bigl\{i\Lambda,W_\alpha\bigr\}
   -\overline D_{\dot\alpha}
   \bigl\{e^{-V}i\Lambda^\dagger e^V,
   \overline W^{\prime\dot\alpha}\bigr\}
   \Bigr),
\cr
}
\eqn\fifteen
$$
which satisfies Eq.~\ten\ in conjunction with~Eq.~\eight; this fact
provides the consistency check of~Eq.~\fourteen. Finally,
{}from~Eqs.~\seven, \eight\ and~\fourteen, we obtain the consistent
gauge anomaly
$$
\eqalign{
   &\delta_\Lambda{\mit\Gamma}[V]
\cr
   &\Mto-{1\over64\pi^2}\int d^6z\,\tr i\Lambda
   W^\alpha W_\alpha
   +{1\over64\pi^2}\int d^6\overline z\,\tr e^{-V}i\Lambda^\dagger e^V
   \overline W_{\dot\alpha}^\prime\overline W^{\prime\dot\alpha}
\cr
   &\qquad
   +{1\over64\pi^2}\int d^8z\,
   \int_0^1dg\int_0^1d\beta\,
   \tr e^{-\beta gV}\delta_\Lambda Ve^{\beta gV}
\cr
   &\qquad\qquad\qquad\qquad\quad
   \times
   \Bigl(
   \bigl[{\cal D}^\alpha V,W_\alpha\bigr]
   +\bigl[\overline D_{\dot\alpha}V,
   \overline W^{\prime\dot\alpha}\bigr]
   +\bigl\{V,{\cal D}^\alpha W_\alpha\bigr\}\Bigr)_{V\to gV}.
\cr
}
\eqn\sixteen
$$
Here, as indicated, the quantities inside the round bracket are
defined by substituting the gauge superfield~$V$ involved by~$gV$.
On the other hand, the gauge variation~$\delta_\Lambda V$ is given
by~Eq.~\five\ as it stands {\it without\/} setting~$V\to gV$. It is
obvious that our consistent anomaly is proportional to the
anomaly~$d^{abc}$, as expected.

As the simple but non-trivial check of Eq.~\sixteen, we may consider
the Abelian case for which the expression is considerably simplified.
By noting~$\delta_\Lambda V=i\Lambda-i\Lambda^\dagger$ in this case,
we have $\delta_\Lambda{\mit\Gamma}[V]=(-1+2/3)%
\int d^6z\,\tr i\Lambda W^\alpha W_\alpha/\allowbreak(64\pi^2)%
+{\rm h.c.}$ This is one-third the covariant anomaly~\eight\ and
reproduces the correct result of the consistent Abelian anomaly.
We note that, in our treatment, nothing special (except simplicity of
the expression) occurs in the Abelian case. In approaches based on
the Wess-Zumino consistency condition, strictly speaking, it is
necessary to start with the non-Abelian case and then to take the
Abelian limit, because the consistency condition becomes trivial in
the Abelian case.

One might ask whether the anomaly~\sixteen\ actually satisfies the
Wess-Zumino consistency condition. In fact, Eq.~\sixteen\ is identical
to the consistent anomaly due to Marinkovi\'c~[\MAR] up to the
overall normalization factor (ours is four times smaller). See Eq.~(5)
of~Ref.~[\MAR], where $g$ is denoted as~$t$
and~$\int_0^1d\beta\,e^{-\beta gV}g\delta_\Lambda Ve^{\beta gV}$ is
abbreviated as~$S_g$. Since the consistent anomaly in~Ref.~[\MAR] was
constructed as the solution of the consistency condition, we may
claim that we already know Eq.~\sixteen\ actually satisfies the
consistency condition.

It is interesting to examine the form of the anomaly~\sixteen\ in the
Wess-Zumino (WZ) gauge $V=-\theta\sigma^m\overline\theta+%
i\theta^2\overline\theta\overline\lambda-
i\overline\theta^2\theta\lambda+\theta^2\overline\theta^2D/2$~[\WESS].
We first set $\Lambda(z)=a(y)$ for reproducing the usual gauge
transformation ($a$~is real). Then we have
$$
\eqalign{
   &\delta_\Lambda{\mit\Gamma}[V]
\cr
   &\Mto-{1\over96\pi^2}\int d^4x\,\tr a
   \biggl[\varepsilon^{mnkl}\partial_m
   \Bigl(v_n\partial_kv_l+{i\over4}v_nv_kv_l\Bigr)
   -{1\over2}\partial_m(\overline\lambda\overline\sigma^m\lambda
   -\lambda\sigma^m\overline\lambda)\biggr].
\cr
}
\eqn\seventeen
$$
This expression of the usual gauge anomaly in the WZ gauge is
surprisingly simple compared to the result of existing field
theoretical calculations. We emphasize again that we obtained
Eq.~\seventeen\ without supplementing any counterterms. The first
term is celebrated Bardeen's minimal anomaly~[\BAR] with the
coefficient for a single chiral fermion. The second term, if one
wishes, may be eliminated by adding a non-supersymmetric local
counterterm~$C$ as~$\delta_\Lambda{\mit\Gamma}[V]+%
\delta_aC$, where $\delta_a$ is the usual gauge transformation
$\delta_av_m=2{\cal D}_ma$ and~$\delta_a\lambda=-i[a,\lambda]$. The
counterterm is given by~[\MAR]
$$
   C={1\over384\pi^2}\int d^4x\,\tr
   v^m(\overline\lambda\overline\sigma_m\lambda
       -\lambda\sigma_m\overline\lambda).
\eqn\eighteen
$$

As another interesting application, we may consider the anomalous
breaking of the supersymmetry in the WZ gauge, the so-called
$\varepsilon$-SUSY anomaly~[\ITO,\GUAD]. The super-transformation in
the WZ gauge is a combination of the supersymmetric transformation
generated by~$\varepsilon Q+\overline\varepsilon\overline Q$
(which is not anomalous in the present formulation~[\HAY]) {\it and\/}
the gauge transformation~$\delta_\Lambda$ with the gauge
parameter~$\Lambda(z)=-i\theta\sigma^m\overline\varepsilon v_m(y)-%
\theta^2\overline\varepsilon\overline\lambda(y)$~[\WESS]. Therefore
we have the (apparent) breaking of supersymmetry as the consequence
of the gauge anomaly. By setting the gauge parameter~$\Lambda$ to
this form in~Eq.~\sixteen, we have after some calculation,
$$
\eqalign{
   &\delta_\Lambda{\mit\Gamma}[V]
\cr
   &\Mto{i\over384\pi^2}\int d^4x\,
   \tr(\overline\varepsilon\overline\sigma^m\lambda
   -\overline\lambda\overline\sigma^m\varepsilon)
   \biggl\{
   3\overline\lambda\overline\sigma_m\lambda
   -\varepsilon_m{}^{nkl}\Bigl[2v_n(\partial_kv_l)+2(\partial_nv_k)v_l
   +{3i\over2}v_nv_kv_l\Bigr]\biggr\}
\cr
   &\qquad\quad-\delta_\varepsilon C,
}
\eqn\nineteen
$$
where $\delta_\varepsilon$~is the super-transformation in the WZ
gauge $\delta_\varepsilon v^m=%
i\overline\varepsilon\overline\sigma^m\lambda+{\rm h.c.}$,
$\delta_\varepsilon\lambda=%
\sigma^{mn}\varepsilon v_{mn}+i\varepsilon D$ and
$\delta_\varepsilon D=%
-{\cal D}_m\lambda\sigma^m\overline\varepsilon+{\rm h.c.}$
Eq.~\nineteen\ shows that Eq.~\sixteen\ reproduces the
$\varepsilon$-SUSY anomaly given in~Ref.~[\GUAD] again with the
non-supersymmetric local counterterm~$C$~\eighteen. Note that the
structure of the counterterm~\eighteen\ is quite simple, compared to
that of the counterterm required in~Ref.~[\GUAD] for obtaining the
above form. Our anomaly is proportional to~$d^{abc}$ from the
beginning and thus the possible (non-supersymmetric) counterterm also
must be proportional to~$d^{abc}$. This fact severely restricts the
possible form of counterterms.

In this letter, we have presented a (yet another) field theoretical
calculation of the consistent gauge anomaly in the superfield
formalism. As we have shown, it is possible to fully utilize the
advantage of gauge covariance, by defining the effective action
through the covariant gauge current~(Eqs.~\two\ and~\three). Although
our result~\sixteen\ itself has been known in the literature~[\MAR]
(see also~Ref.~[\MCA]), this is the first time to our knowledge that
an explicit field theoretical calculation in the superfield formalism
directly led to the minimal consistent anomaly.


\Appendix{A}
\noindent
Notational conventions:
$$
\eqalign{
   &\eta_{mn}=\mathop{\rm diag}(-1,+1,+1,+1),
\cr
   &z=(x^m,\theta^\alpha,\overline\theta_{\dot\alpha}),\quad
   y^m=x^m+i\theta\sigma^m\overline\theta,
\cr
   &d^8z=d^4x\,d^2\theta\,d^2\overline\theta,\quad
   d^6z=d^4x\,d^2\theta,\quad
   d^6\overline z=d^4x\,d^2\overline\theta,
\cr
   &\delta(z)=\delta(x)\delta(\theta)\delta(\overline\theta),
\cr
   &W_\alpha=-{1\over4}\overline D^2(e^{-V}D_\alpha e^V),\quad
   \overline W_{\dot\alpha}^\prime
   ={1\over4}e^{-V}D^2(e^V\overline D_{\dot\alpha}e^{-V})e^V,
\cr
   &\nabla_\alpha=e^{-V}D_\alpha e^V,\quad
   \bigl\{\nabla_\alpha,\overline D_{\dot\alpha}\bigr\}
   =-2i\sigma^m_{\alpha\dot\alpha}\nabla_m,
\cr
   &{\cal D}^\alpha A=\bigl\{\nabla^\alpha,A\,\bigr],
\cr
   &\dA_+
   ={1\over16}\overline D^2\nabla^2+{1\over16}\nabla^2\overline D^2
   -{1\over8}\overline D_{\dot\alpha}\nabla^2\overline D^{\dot\alpha}
   =\nabla^m\nabla_m-{1\over2}W^\alpha\nabla_\alpha
   -{1\over4}({\cal D}^\alpha W_\alpha),
\cr
   &\dA_-
   ={1\over16}\nabla^2\overline D^2+{1\over16}\overline D^2\nabla^2
   -{1\over8}\nabla^\alpha\overline D^2\nabla_\alpha
   =\nabla^m\nabla_m
   +{1\over2}\overline W_{\dot\alpha}^\prime\overline D^{\dot\alpha}
   +{1\over4}(\overline D_{\dot\alpha}\overline W^{\prime\dot\alpha}).
\cr
}
\eqn\aone
$$
Identities:
$$
\eqalignno{
   &{\cal D}^\alpha W_\alpha
   =\overline D_{\dot\alpha}\overline W^{\prime\dot\alpha}.
   &\eqnalign{\atwo}
\cr
   &\overline D^2\dA_+=\dA_+\overline D^2
   ={1\over16}\overline D^2\nabla^2\overline D^2.
   &\eqnalign{\athree}
\cr
   &\nabla^2\dA_-=\dA_-\nabla^2
   ={1\over16}\nabla^2\overline D^2\nabla^2.
   &\eqnalign{\afour}
\cr
   &D_\alpha\tr(AB)=\tr{\cal D}_\alpha AB\pm\tr A{\cal D}_\alpha B.
   &\eqnalign{\afive}
\cr
}
$$
Note that the last identity~\afive\ allows the integration by parts
on~${\cal D}_\alpha$.

\Appendix{B}
In this appendix, we illustrate the calculation of~Eq.~\thirteen\ in
plane wave basis. Basically the same calculation was performed
in~Ref.~[\MCA] by using the heat kernel expansion. First,
in~Eq.~\thirteen, we
note~$\delta(x-x')=M^4\int d^4k\,e^{iMk(x-x')}/(2\pi)^4$ and
$$
\eqalign{
   &e^{-iMkx}\nabla_me^{iMkx}
   =\nabla_m+iMk_m,
\cr
   &e^{-iMkx}\nabla_\alpha e^{iMkx}
   =\nabla_\alpha-\sigma_{\alpha\dot\alpha}^m
   \overline\theta^{\dot\alpha}Mk_m,
\cr
   &e^{-iMkx}\overline D_{\dot\alpha}e^{iMkx}
   =\overline D_{\dot\alpha}
   +\theta^\alpha\sigma_{\alpha\dot\alpha}^mMk_m.
\cr
}
\eqn\bone
$$
Then we have
$$
\eqalign{
   &\delta_1\VEV{\delta_2S}-\delta_2\VEV{\delta_1S}
\cr
   &=-{i\over16}\int d^8z\,\int_0^1d\beta\,M^2
   \int{d^4k\over(2\pi)^4}
\cr
   &\quad\times\tr\Delta_2
   \exp\biggl\{-\beta\Bigl[k^mk_m-2ik^m\nabla_m/M
\cr
   &\qquad\qquad\qquad\quad
   -\nabla^m\nabla_m/M^2+W^\alpha\nabla_\alpha/(2M^2)
   +({\cal D}^\alpha W_\alpha)/(4M^2)\Bigr]\biggr\}\overline D^2
\cr
   &\quad\times
   \Delta_1\exp\biggl\{-(1-\beta)\Bigl[k^mk_m-2ik^m\nabla_m/M
\cr
   &\qquad\qquad\qquad\quad
   -\nabla^m\nabla_m/M^2
   -\overline W_{\dot\alpha}^\prime\overline D^{\dot\alpha}/(2M^2)
   -(\overline D_{\dot\alpha}
   \overline W_{\dot\alpha}^\prime)/(4M^2)\Bigr]\biggr\}
   \nabla^2
\cr
   &\qquad\qquad\qquad\qquad\qquad\qquad\qquad
   \times
   \delta(\theta-\theta')
   \delta(\overline\theta-\overline\theta')
   \Bigr|_{\theta=\theta',\overline\theta=\overline\theta'}
   -(1\leftrightarrow2).
\cr
}
\eqn\btwo
$$
In writing this expression, we have omitted terms in which
$\theta^\alpha$ or~$\overline\theta_{\dot\alpha}$ explicitly appears;
the reason for this is the following. The superfield, such
as~Eq.~\thirteen, cannot have a term which explicitly contains
$\theta^\alpha$ or~$\overline\theta_{\dot\alpha}$, because such a
term has no first ($\theta=\overline\theta=0$) component and thus has
no higher components (the higher components of the superfield are
uniquely determined by the linearly realized super-transformation of
the first component~[\WESS]). Therefore those terms in which
$\theta^\alpha$ or~$\overline\theta_{\dot\alpha}$ explicitly appears
must eventually be canceled out, or, if these contribute,
$\theta^\alpha$ must be eliminated by~$D_\alpha$ (or
$\overline\theta_{\dot\alpha}$ by $\overline D^{\dot\alpha}$).
However, in the original form of~\btwo, one can confirm that when
$D_\alpha$ eliminates $\theta^\alpha$ (or when
$\overline D^{\dot\alpha}$ eliminates
$\overline\theta_{\dot\alpha}$), the corresponding term does not have
enough powers of~$M$ to survive in the $M\to\infty$~limit, or does
not have a sufficient number of spinor derivatives to eliminate the
delta function: In the equal point limit~$\theta=\theta'$
and~$\overline\theta=\overline\theta'$, only those terms with which
just four spinor derivatives acting on the delta function
survive~[\WESS]:
$$
   \nabla_\alpha\nabla_\beta\left.\delta(\theta-\theta')
   \right|_{\theta=\theta'}=-2\varepsilon_{\alpha\beta},
   \qquad
   \overline D_{\dot\alpha}\overline D_{\dot\beta}
   \left.\delta(\overline\theta-\overline\theta')
   \right|_{\overline\theta=\overline\theta'}
   =2\varepsilon_{\dot\alpha\dot\beta}.
\eqn\bthree
$$
In short, the terms in which $\theta^\alpha$
or~$\overline\theta_{\dot\alpha}$ explicitly appears must be
canceled out. This cancellation may directly be verified as was
done in Ref.~[\HAY] in a similar calculation.

The expansion of Eq.~\btwo\ in powers of~$1/M$ is easy. After the
integration over~$k_m$ and~$\beta$, one can readily verify (say, by
substituting~$\nabla_m=\partial_m+{\mit\Gamma}_m$) that the terms
contain the vector covariant derivative~$\nabla_m$ are combined into
a total divergence. Thus vector covariant derivatives do not
contribute. In this way, we have
$$
\eqalign{
   &\delta_1\VEV{\delta_2S}-\delta_2\VEV{\delta_1S}
\cr
   &\Mto-{1\over64\pi^2}\int d^8z\,
   \tr\biggl[
   \Delta_2W^\alpha\nabla_\alpha\overline D^2\Delta_1\nabla^2
   +{1\over2}
   \Delta_2({\cal D}^\alpha W_\alpha)\overline D^2\Delta_1\nabla^2
\cr
   &\qquad\qquad\qquad\qquad\qquad\quad
   -\Delta_2\overline D^2\Delta_1
   \overline W_{\dot\alpha}^\prime\overline D^{\dot\alpha}\nabla^2
   -{1\over2}\Delta_2\overline D^2\Delta_1
   (\overline D_{\dot\alpha}
   \overline W^{\prime\dot\alpha})\nabla^2\biggr]
\cr
   &\qquad\qquad\qquad\qquad\qquad\qquad
   \times{1\over16}
   \delta(\theta-\theta')
   \delta(\overline\theta-\overline\theta')
   \Bigr|_{\theta=\theta',\overline\theta=\overline\theta'}
   -(1\leftrightarrow2).
\cr
}
\eqn\bfour
$$
By noting again that four spinor derivatives have to act on the delta
function as in Eq.~\bthree, we find
$$
\eqalign{
   &\delta_1\VEV{\delta_2S}-\delta_2\VEV{\delta_1S}
\cr
   &\Mto-{1\over64\pi^2}\int d^8z
\cr
   &\quad\times\tr\biggl[
   \Delta_2W^\alpha{\cal D}_\alpha\Delta_1
   +{1\over2}\Delta_2({\cal D}^\alpha W_\alpha)\Delta_1
   +\Delta_2\overline D_{\dot\alpha}
   (\Delta_1\overline W^{\prime\dot\alpha})
   -{1\over2}\Delta_2\Delta_1
   \overline D_{\dot\alpha}\overline W^{\prime\dot\alpha}\biggr]
\cr
   &\qquad\qquad\qquad\qquad\qquad\qquad\qquad\qquad
   \qquad\qquad\qquad\qquad\qquad\quad
   -(1\leftrightarrow2).
\cr
}
\eqn\bfive
$$
Finally, we obtain Eq.~\fourteen\ after some rearrangements with use
of the reality constraint~\atwo\ and the identity~\afive.

\refout
\bye